\documentclass{appolb}
\usepackage{graphicx,amssymb,epsfig,epsf}
\usepackage{amsmath}
\usepackage{axodraw}
\begin{document}
\title{The MSSM with large $\tan\beta$ beyond the decoupling limit%
\thanks{Presented at FLAVIAnet Topical Workshop ``Low energy constraints on extensions
of the Standard Model'', Kazimierz, Poland, 23-27 July 2009.}%
}
\author{LARS HOFER\thanks{in collaboration with Ulrich Nierste and Dominik Scherer}
\address{Institut f\"ur Theoretische Physik, Karlsruhe Institute of Technology,\\
         D-76128 Karlsruhe, Germany}
}
\maketitle
\begin{abstract}
  For large values of $\tan\beta$ interesting effects arise in the MSSM due to the enhancement of down-quark self-energies. These effects are well-studied within the decoupling limit, i.e. in the limit of supersymmetric masses far above the electroweak scale. In this article I discuss those issues which emerge from a treatment of these effects that goes beyond the decoupling limit. 
\end{abstract}
  
\section{Introduction}
The Minimal Supersymmetric Standard Model (MSSM) contains two Higgs doublets
$H_u$ and $H_d$ coupling to up- and down-type quark fields, respectively. The neutral components of these Higgs doublets acquire vacuum expectation values (vevs) $v_u$ and $v_d$ with the sum $v_u^2+v_d^2$ being fixed to $v^2\approx
(174\,\,\mbox{GeV})^2$ and the ratio $\tan\beta\equiv v_u/v_d$ remaining as a free parameter. Large values of $\tan\beta$ ($\sim50$) are theoretically motivated by bottom-top Yukawa unification, which occurs in SO(10) GUT models with
minimal Yukawa sector, and phenomenologically preferred by the anomalous magnetic moment of
the muon \cite{Bennet:2006}. Since a large value of $\tan\beta$ corresponds to $v_d\ll v_u$, it leads to enhanced corrections in amplitudes where the tree-level contribution is suppressed by the small vev $v_d$ but the loop-correction involves $v_u$ instead. In such cases the ratio of one-loop to tree-level contribution receives an enhancement-factor $\tan\beta$ which may lift the loop-suppression rendering the ratio of order $\mathcal{O}(1)$. 

In order not to spoil the perturbative expansion a special treatment is required for these $\tan\beta$-enhanced corrections to resum them to all orders. There are two possible ways to deal with them. The first is to consider an effective theory with the SUSY particles intgrated out keeping only Higgs fields and SM particles \cite{YbMb,CKM,NeuH}. This approach is valid for $M_{\textrm{SUSY}}\gg v,\,M_{A^0,H^0,H^{\pm}}$, i.e. within the decoupling limit; it can be extended beyond using an iterative method \cite{decoupit}. The second possibility is to perform a diagrammatic, analytic resummation in the full MSSM without assuming any hierarchy between $M_{\textrm{SUSY}}$, $M_{A^0,H^0,H^{\pm}}$ and $v$ \cite{CGNW,HNS}. Going beyond the decoupling limit is desirable since on the one hand $M_{\textrm{SUSY}}\sim v$ is natural and on the other hand $\tan\beta$-enhanced effects in couplings involving SUSY particles like gluinos and neutralinos cannot be studied in an effective theory with these particles integrated out. In this article I will present the main results of Ref. \cite{HNS}, in which the diagrammatic resummation method of Ref. \cite{CGNW} has been extended to the case of flavour-changing interactions.

\section{Scheme dependence of the resummation formula for the Yukawa coupling $y_b$} 
At tree-level the mass of the bottom quark $m_b$ is generated by coupling the $b$-quark to the Higgs-field $H_d$ and it is thus proportional to the small vev $v_d$. At one-loop it receives $\tan\beta$-enhanced self-energy corrections (involving $v_u$ instead of $v_d$) from gluino-sbottom-, chargino-stop- and neutralino-sbottom-loops, which I parameterise as
\begin{equation}
   \Sigma^{RL}_b\,\,=\,\,m_b\,\Delta_b\,\,=\,\,m_b\,\epsilon_b\,\tan\beta.
   \label{eq:FlavConsSelfEn}
\end{equation}

These corrections modify the relation between the Yukawa-coupling $y_b$ and the $\overline{\textrm{MS}}$ mass $m_b$. Within the decoupling limit the resulting resummation formula is given by \cite{YbMb}
\begin{equation}
   y_b\,\,=\,\,\frac{m_b}{v_d\,(1\,+\,\Delta_b)}.
\end{equation}

Beyond the decoupling limit the form of the resummation formula actually depends on the choice of renormalisation scheme for the input parameters in the sbottom sector. Neclecting $\tan\beta$-suppressed terms, the sbottom mass matrix reads
\begin{equation}
   \mathcal{M}^2_{\tilde{b}}\,\,=\,\,\begin{pmatrix} m_{\tilde{b}_L}^2 & -y_b^*\,v_u\,\mu \\ 
                                             -y_b\,v_u\,\mu^* & m_{\tilde{b}_R}^2  \end{pmatrix}.
\end{equation}
It is diagonalised by a mixing matrix $\widetilde{R}_b$ involving a mixing angle $\tilde{\theta}_b$ and a phase $\tilde{\phi}_b$:
\begin{equation}
   \widetilde{R}_b\,\mathcal{M}^2_{\tilde{b}}\,\widetilde{R}^{\dagger}_b\,\,=\,\,
   \textrm{diag}(m_{\tilde{b}_1}^2,\,m_{\tilde{b}_2}^2),\hspace{1cm}
   \widetilde{R}_b\,\,=\,\,\begin{pmatrix} \cos\tilde{\theta}_b & \sin\tilde{\theta}_b\,e^{i\tilde{\phi}_b} \\
                                          -\sin\tilde{\theta}_b\,e^{-i\tilde{\phi}_b} & \cos\tilde{\theta}_b
                           \end{pmatrix}.
\end{equation}

If one takes the elements of the mass matrix $\mathcal{M}_{\tilde{b}}^2$ as input, the paramters $m_{\tilde{b}_{1,2}}$, $\tilde{\theta}_b$ and $\tilde{\phi}_b$ are fixed by the diagonalisation procedure, i.e. they are not free parameters but functions of the elements of $\mathcal{M}_{\tilde{b}}^2$. On the other hand, can also take $m_{\tilde{b}_{1,2}}$, $\tilde{\theta}_b$ and $\tilde{\phi}_b$ (or other combinations of parameters) directly as input. 
Note that it is not possible to distinguish between different schemes for the input parameteres in the limit $v/M_{\textrm{SUSY}}\to 0$ because
\begin{equation}
    m^2_{\tilde{b}_{1,2}}=m^2_{\tilde{b}_{L,R}}\,\left(1+\mathcal{O}\left(v^2/M_{\textrm{SUSY}}^2\right)\right),\hspace{1cm}
    \sin 2\tilde{\theta}_b=\mathcal{O}\left(v/M_{\textrm{SUSY}}\right).
\end{equation}
Beyond the decoupling limit, however, different choices for the set of input parameters lead to different resummation formulae for $y_b$, which are shown in table \ref{tab:SchemeDep}. Here, only the gluino-contribution $\Delta_b^{\tilde{g}}$ of $\Delta_b$ is considered; formulae including also the chargino- and the neutralino-contribution as well as explicit expressions for $\Delta_b^{\tilde{g}}$, $\Delta_b^{\widetilde{\chi}^{\pm}}$ and $\Delta_b^{\widetilde{\chi}^0}$ can be found in Ref. \cite{HNS}.
\begin{table}[t!] 
  \begin{center}
  \begin{tabular}{|l|l|}
     \hline &\vspace{-0.2cm}\\
     \normalsize{input} &
         \normalsize{resummation formula} \vspace{-0.2cm}\\ &\\
     \hline\hline &
                    \vspace{-0.2cm}\\
                    $m_{\tilde b_{1}}$, $m_{\tilde b_{2}}$,
                    $\tilde{\theta}_b$, $\tilde{\phi}_b$ &
                    \hspace{0.2cm}$y_b=\dfrac{m_b}{v_d}\left(1-\Delta_b^{\tilde{g}}\right)$\vspace{-0.2cm}\\ 
                    &
                    \vspace{-0.1cm}\\
     \hline &\vspace{-0.2cm}\\
                    $m_{\tilde b_{1}}$, $m_{\tilde b_{2}}$, $\mu$, $\tan\beta$ &
                    \hspace{0.2cm}$y_b=\dfrac{m_b}{v_d(1+\Delta_b^{\tilde{g}})}$\vspace{-0.2cm}\\ &\vspace{-0.1cm}\\
     \hline &\vspace{-0.2cm}\\
                    $m_{\tilde b_{L}}$, 
                    $m_{\tilde b_{R}}$, $\mu$, $\tan\beta$  &
                    $\begin{array}{l} \textrm{analytic resummation impossible,} \\ 
                    \textrm{use formula (i) iteratively.}\end{array}$\vspace{-0.2cm}\\
     &\\
     \hline 
  \end{tabular}
  \end{center}
  \caption{resummation formulae for $y_b$ for different choices of the input parameters}
  \label{tab:SchemeDep}
\end{table}

\section{Effects of flavour-changing self-energies}
In the last section I briefly discussed the consequences of $\tan\beta$-enhanced contributions to the $b$-quark self-energy $\Sigma^{RL}_{b}$. Let us now have a look at the corresponding flavour-changing self-energies and let us focus on $b\to s$ transitions $\Sigma^{RL}_{bs}$ for definiteness. In the framework of naive MFV as defined in Ref. \cite{HNS}, which is considered here, the gluino- and neutralino-couplings are flavour-diagonal at tree-level. Therefore only chargino diagrams generate a $\tan\beta$-enhanced contribution to $\Sigma^{RL}_{bs}$ which can be parameterised analogously to Eq.
(\ref{eq:FlavConsSelfEn}) as
\begin{equation}
   \Sigma^{RL}_{bs}\,\,=\,\,V_{tb}^*\,V_{ts}\,m_b\,\Delta_{FC}
                   \,\,=\,\,V_{tb}^*\,V_{ts}\,m_b\,\epsilon_{FC}\,\tan\beta.
\end{equation}

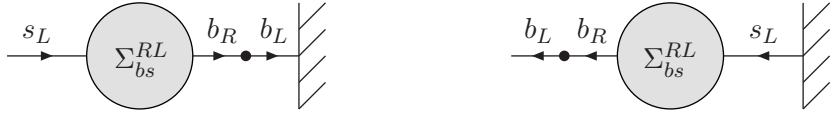
\begin{figure}[t]
  \begin{picture}(200,35) (-200,-20)
    \SetWidth{0.5}
    \SetColor{Black}
    \ArrowLine(-170,0)(-140,0)
    \ArrowLine(-100,0)(-80,0)
    \ArrowLine(-80,0)(-60,0)
    \Vertex(-80,0){2}
    \GOval(-120,0)(20,20)(0){0.882}
    \Line(-60,-20)(-60,20)
    \Line(-60,10)(-50,20)
    \Line(-60,0)(-50,10)
    \Line(-60,-10)(-50,0)
    \Line(-60,-20)(-50,-10)
    \Text(-165,5)[lb]{\Black{$s_L$}}
    \Text(-95,5)[lb]{\Black{$b_R$}}
    \Text(-75,5)[lb]{\Black{$b_L$}}
    \Text(-130,-8)[lb]{\Black{$\Sigma^{RL}_{bs}$}}
   % \Text(-120,-40)[lb]{(1)}

    \ArrowLine(40,0)(20,0)
    \ArrowLine(60,0)(40,0)
    \ArrowLine(130,0)(100,0)
    \Vertex(40,0){2}
    \GOval(80,0)(20,20)(0){0.882}
    \Line(130,-20)(130,20)
    \Line(130,10)(140,20)
    \Line(130,0)(140,10)
    \Line(130,-10)(140,0)
    \Line(130,-20)(140,-10)
    \Text(25,5)[lb]{\Black{$b_L$}}
    \Text(45,5)[lb]{\Black{$b_R$}}
    \Text(110,5)[lb]{\Black{$s_L$}}
    \Text(70,-8)[lb]{\Black{${\Sigma^{RL}_{bs}}$}}
   % \Text(75,-40)[lb]{(2)}
  \end{picture}
\caption{Feynman diagrams with $\Sigma^{RL}_{bs}$ in  
 an external quark leg.}
\label{fig:ExtLegSelf}
\end{figure}

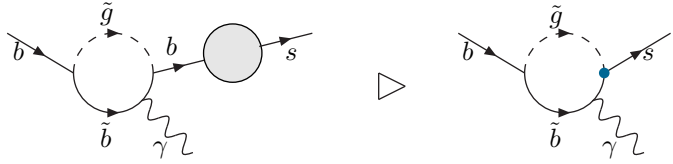
\begin{figure}[t]
     \begin{picture}(200,55)(-70,-15)
        \SetWidth{0.5}
        \SetColor{Black}
        \SetWidth{0.5}
        \ArrowLine(150,30)(175,15)
        \ArrowArc(190,15)(15,180,0)
        \DashArrowArcn(190,15)(15,180,0){3}
        \ArrowLine(205,15)(230,30)
        \Photon(200.5,4.5)(220,-15){3}{3}
        \Text(152,20)[lb]{\small{$b$}}
        \Text(220,20)[lb]{\small{$s$}}
        \Text(185,-13)[lb]{\small{$\tilde{b}$}}
        \Text(185,32)[lb]{\small{$\tilde{g}$}}
        \Text(205,-17)[lb]{\small{$\gamma$}}

        \Line(120,5)(120,15)
        \Line(120,5)(130,10)
        \Line(120,15)(130,10)

        \ArrowLine(-20,30)(5,15)
        \ArrowArc(20,15)(15,180,0)
        \DashArrowArcn(20,15)(15,180,0){3}
        \ArrowLine(35,15)(55,20)
        \GOval(65,22.5)(11,11)(0){0.9}
        \ArrowLine(75,25)(95,30)
        \Photon(30.5,4.5)(50,-15){3}{3}
        \Text(-18,20)[lb]{\small{$b$}}
        \Text(85,20)[lb]{\small{$s$}}
        \Text(40,22)[lb]{\small{$b$}}
        \Text(15,-13)[lb]{\small{$\tilde{b}$}}
        \Text(15,33)[lb]{\small{$\tilde{g}$}}
        \Text(35,-17)[lb]{\small{$\gamma$}}
        \SetColor{MidnightBlue}
        \Vertex(205,15){2}
     \end{picture}
     \caption{Gluino diagram giving a contribution to $b\to s\gamma$ of the form $(\textrm{loop}\times\tan\beta)^2$. Left: External leg method. Right: Effective vertex method.}
     \label{fig:gluonC7}
  \end{figure}

Now consider the generic situation of a $\Sigma^{RL}_{bs}$-subdiagram in an external quark leg of some Feynman diagram (fig. \ref{fig:ExtLegSelf}). Since in such a diagram the quark propagator $\pm i/m_b$ ($m_s$ is set to zero) cancels the factor $m_b$ in $\Sigma^{RL}_{bs}$, the resulting Feynman amplitude is given by
\begin{equation}
   \mathcal{M}\,\,=\,\,\pm\,\mathcal{M}^{\textrm{rest}}\cdot V_{tb}^*\,V_{ts}\,\epsilon_{FC}\,\tan\beta
   \label{eq:ExtLegCor}
\end{equation}
where $\mathcal{M}^{\textrm{rest}}$ stands for the part of $\mathcal{M}$ corresponding to the truncated diagram. 
Therefore, if $\tan\beta$ is large enough to compensate for the loop-factor $\epsilon_{FC}$, it is possible to get a $b\to s$ transition without paying the price of a loop suppression. In this way one finds $\tan\beta$-enhanced corrections to FCNC $b\to s$ processes by sourcing the flavour change out into an external quark leg rendering the underlying loop-diagram flavour-diagonal. As an example consider the SUSY-contribution to $b\to s\gamma$: For large $\tan\beta$ the one-loop amplitude is dominated by the chargino-stop diagram giving a contribution of the form $(\textrm{loop}\times\tan\beta)$. Taking into account the above mentioned external leg corrections one finds contributions involving gluino-sbottom loops, like the left diagram in fig. \ref{fig:gluonC7}, which are of the form $(\textrm{loop}\times\tan\beta)^2$.

Alternatively, since $p^2\ll M_{\textrm{SUSY}}^2$ for a momentum $p$ of an on-shell quark and since thus the external leg corrections are local, it is possible to promote their effects to effective flavour-changing vertices, even beyond the decoupling limit because no assumption about a hierarchy between $M_{\textrm{SUSY}}$ and the electroweak scale $v$ is needed. In addition to the FCNC couplings of neutral Higgs bosons, which are well-studied within the decoupling limit \cite{NeuH}, also FCNC couplings of the gluino and neutralino arise in this way. A complete set of Feynman rules for these vertices, including contributions of the form $(\textrm{loop}\times\tan\beta)^n$ to all orders $n=1,2,...$, is given in Ref. \cite{HNS}. The $\tan\beta$-enhanced gluino-sbottom contribution to $b\to s\gamma$ discussed above is within this framework then simply given by a one-loop diagram involving the new FCNC gluino coupling (right digram in fig.
\ref{fig:gluonC7}). 

\section{Sizable gluino contribution to $C_8$}
The effects discussed in the last section give rise to new contributions to the Wilson coefficients of the effective $\Delta B=1$ and $\Delta B=2$ Hamiltonians. Most of these contributions turn out to be numerically small for two reasons: Firstly, for positive $\mu$ and $\tan\beta\sim 50$ typical values $\Delta_{FC}$ are $\sim\, 0.12\ll 1$. Secondly, the gluino contributions, which should give the largest effects due to the strong coupling constant $g_s$, suffer from a GIM suppression.  
\begin{figure}[b]
    \begin{minipage}[t]{0.35\linewidth}
        \includegraphics[width=\textwidth]{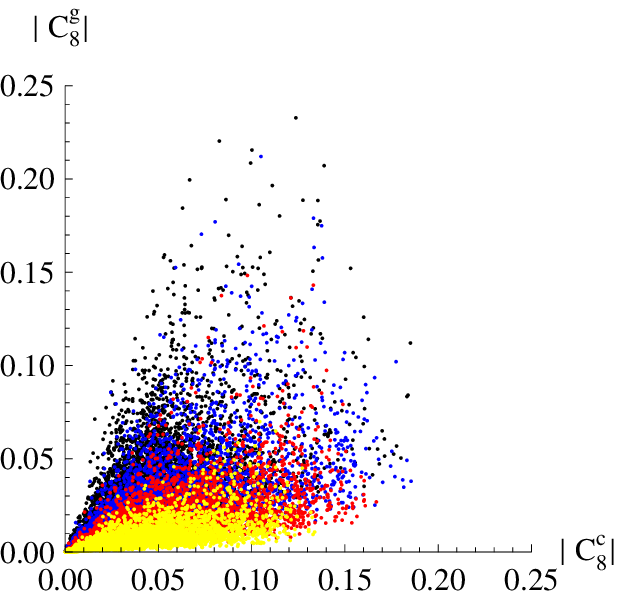}
    \end{minipage}\hspace{0.1\linewidth}
    \begin{minipage}[t]{0.55\linewidth}
         \includegraphics[width=\textwidth]{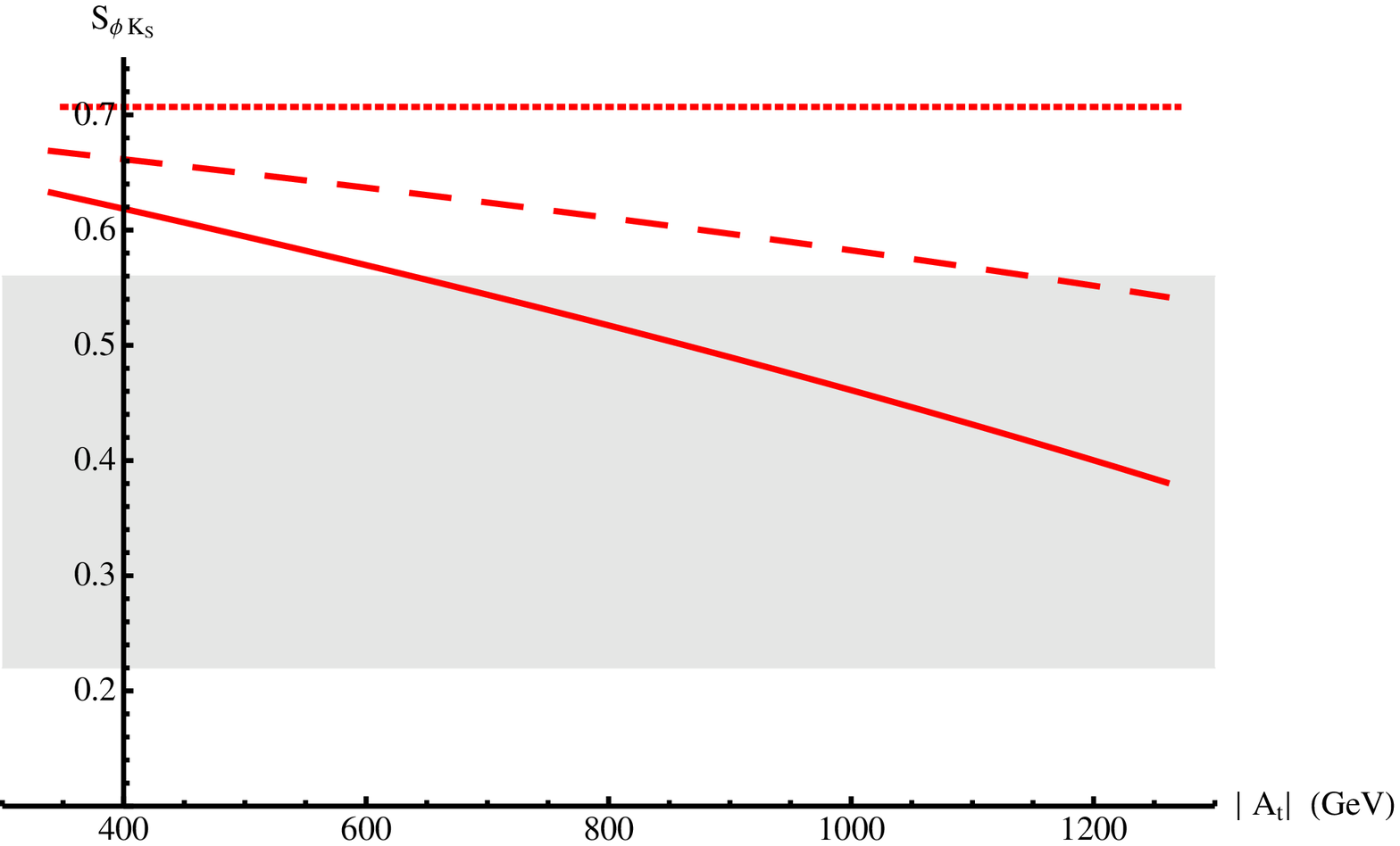}
    \end{minipage}
\caption{Left: Magnitudes of chargino and gluino contributions to $C_8$ scanned over the MSSM parameter space. Right: $S_{\phi K_S}$ as a funtion of $|A_t|$.}
\label{fig:plots}
\end{figure}

There is one exception: Chirally enhanced contributions to the magnetic and chromomagnetic operators $\mathcal{O}_7$ and $\mathcal{O}_8$ (like the diagram in fig. \ref{fig:gluonC7}) involve a left-right-flip in the squark-line proportional to the corresponding quark mass und thus avoid the GIM cancellation. Whereas the contribution from gluino-squark loops to $C_7$ is accidentally small, the one to $C_8$ can indeed contribute as much as the chargino-squark diagram. This can be seen from the left diagram in fig. \ref{fig:plots} where the magnitudes of both contributions $|C_8^c|$ and $|C_8^g|$ are shown for a scan over the MSSM parameter space with positive $\mu$. The colour code (yellow: $200\,\textrm{GeV}<\mu<\,400\,\textrm{GeV}$, red: $400\,\textrm{GeV}<\mu<600\,\textrm{GeV}$,
blue: $600\,\textrm{GeV}<\mu<800\,\textrm{GeV}$, black: $800\,\textrm{GeV}<\mu<1000\,\textrm{GeV}$) reflects the fact that the importance of $C_8^g$ grows with $\mu$. All points in the plot are in agreement with the constraints from $\mathcal{B}(\bar{B}\to X_s\gamma)$ and the experimental lower bounds for the sparticle and lightest Higgs Boson masses. An arbitrary phase is allowed for the parameter $A_t$, however, to avoid the possibility of fulfilling the $\mathcal{B}(\bar{B}\to X_s\gamma)$ constraint by an unnatural fine-tuning of this phase, the additional condition $|C_7^{\textrm{SUSY}}|<|C_7^{\textrm{SM}}|$ is imposed. 

As a consequence one expects significant effects in those low energy observables with a strong dependence on $C_8$. To illustrate this fact the mixing-induced CP asymmetry $S_{\phi K_S}$ of the decay $\bar{B}^0\to\phi\,K_s$ is plotted as a function of $|A_t|$ in the right diagram of fig.\,\ref{fig:plots}. The parameter point chosen for the plot fulfills all constraints mentioned above. The shaded area represents the experimental $1\sigma$ range, the dotted line the SM contribution in leading-order QCD factorisation. For the results corresponding to the dashed and the solid lines the effects of $C_8^c$ and $C_8^c+C_8^g$ have been taken into account, respectively. The plot demonstrates that for complex $A_t$ the gluino-squark contribution can indeed have a large impact on $S_{\phi K_S}$.%, especially if $|A_t|$ is large.

\section{Conclusions}
The effects of $\tan\beta$-enhanced self-energies can be resummed analytically beyond the decoupling limit, also in the flavour-changing case. The resummation formula for the Yukawa coupling $y_b$ depends on the renormalisation scheme used for the input parameters in the sbottom sector. Moreover not only the neutral Higgs bosons but also the gluino and the neutralino develop FCNC couplings for large values of $\tan\beta$. This leads to a sizable modification of the Wilson coefficient $C_8$ of the chromomagnetic operator.

\subsection*{Acknowledgments}
I am grateful to M. Krawczyk, H. Czyz and M. Misiak for organizing
this very pleasant and stimulating workshop. I thank U. Nierste and D. Scherer for the enjoyable collaboration on the presented work \cite{HNS}.

\end{document}